\documentclass[sigconf,nonacm]{acmart}

\usepackage{graphicx} 
\usepackage{import} 
\usepackage{xspace} 
\usepackage{numprint} 
\usepackage{ifthen}
\usepackage{listings}
\usepackage{subcaption} 
\usepackage{multirow}
\usepackage[nameinlink]{cleveref}
\usepackage{caption}
\usepackage{framed}
\usepackage{tikz}
\usepackage{tikz-qtree}
\usepackage[epsilon,altpo]{backnaur}
\usepackage{bussproofs}
\usepackage{booktabs}
\usepackage{tabularx}
\usepackage{multirow}
\usepackage{pifont}
\usepackage{wasysym}
\usepackage{makecell}
\usepackage{mathtools}
\usepackage{pgfplotstable}
\usepackage{enumitem}
\theoremstyle{plain}
\newtheorem{finding}{Finding}

\newboolean{showcomments}
\setboolean{showcomments}{true} 
\ifthenelse{\boolean{showcomments}}{
  \newcommand{\nbc}[3]{
    {\textcolor{#3}{\small{\bfseries{#1:\ }}\textit{#2}}}}
}{
  \newcommand{\nbc}[3]{}
}

\newlist{questions}{enumerate}{2}
\setlist[questions,1]{label=RQ\arabic*.,ref=RQ\arabic*}
\setlist[questions,2]{label=(\alph*),ref=\thequestionsi(\alph*)}

\setcopyright{none}

\title{User-Centric Deployment of Automated Program Repair at Bloomberg}

\author{David Williams}
\affiliation{%
  \institution{University College London}
  \streetaddress{Gower St, London WC1E 6BT}
  \country{United Kingdom}}
\email{david.williams.22@ucl.ac.uk}

\author{James Callan}
\affiliation{%
  \institution{University College London}
  \streetaddress{Gower St, London WC1E 6BT}
  \country{United Kingdom}}
\email{james.callan.19@ucl.ac.uk}

\author{Serkan Kirbas}
\affiliation{%
  \institution{Bloomberg}
  \streetaddress{3 Queen Victoria St, London EC4N 4TQ}
  \country{United Kingdom}}
\email{skirbas@bloomberg.net}

\author{Sergey Mechtaev}
\affiliation{%
  \institution{University College London}
  \streetaddress{Gower St, London WC1E 6BT}
  \country{United Kingdom}}
\email{s.mechtaev@ucl.ac.uk}

\author{Justyna Petke}
\affiliation{%
  \institution{University College London}
  \streetaddress{Gower St, London WC1E 6BT}
  \country{United Kingdom}}
\email{j.petke@ucl.ac.uk}

\author{Thomas Prideaux-Ghee}
\affiliation{%
  \institution{Bloomberg}
  \streetaddress{3 Queen Victoria St, London EC4N 4TQ}
  \country{United Kingdom}}
\email{tprideauxghe@bloomberg.net}

\author{Federica Sarro}
\affiliation{%
  \institution{University College London}
  \streetaddress{Gower St, London WC1E 6BT}
  \country{United Kingdom}}
\email{f.sarro@ucl.ac.uk}

\begin{document}

\begin{abstract}
  Automated program repair (APR) tools have unlocked the potential for the rapid rectification of codebase issues. However, to encourage wider adoption of program repair in practice, it is necessary to address the usability concerns related to generating irrelevant or out-of-context patches. When software engineers are presented with patches they deem uninteresting or unhelpful, they are burdened with more "noise" in their workflows and become less likely to engage with APR tools in future. This paper presents a novel approach to optimally time, target, and present auto-generated patches to software engineers. To achieve this, we designed, developed, and deployed a new tool dubbed B-Assist, which leverages GitHub's Suggested Changes interface to seamlessly integrate automated suggestions into active pull requests (PRs), as opposed to creating new, potentially distracting PRs. This strategy ensures that suggestions are not only timely, but also contextually relevant and delivered to engineers most familiar with the affected code. Evaluation among Bloomberg software engineers demonstrated their preference for this approach. From our user study, B-Assist's efficacy is evident, with the acceptance rate of patch suggestions being as high as 74.56\%; engineers also found the suggestions valuable, giving usefulness ratings of at least 4 out of 5 in 78.2\% of cases. Further, this paper sheds light on persisting usability challenges in APR and lays the groundwork for enhancing the user experience in future APR tools.
\end{abstract}

\maketitle

\section{Introduction}
\label{sec:intro}
In recent years, the advent of automated program repair (APR) tools has brought with it the promise of swift resolutions to codebase issues. However, with the proliferation of these tools, a new challenge has emerged: determining when and how these auto-generated patches should be presented to software engineers to maximise their efficacy and acceptance rate.

Pull requests (PRs) are a standard format to introduce changes into a codebase. While PRs serve as a gatekeeping mechanism to ensure code quality and maintainability, reviewing PRs demands a significant cognitive load from software engineers, requiring them to shift focus, context-switch, and allocate time to evaluate proposed changes. Moreover, once a PR is accepted and merged, the modified code must be deployed to the production environment. This necessitates additional cognitive load and work for software engineers, so it is crucial that automated patches do not exacerbate the burden but rather seamlessly integrate with their existing workflows.
Thus, opening new PRs with automatically generated patches without due consideration to the relevance of the changes or the expertise of the assigned reviewers, which we refer to as \emph{blind pull requests}, is not a sustainable or effective strategy.

Previous studies on the deployment of APR in industry reported the acceptance rate of generated patches to be around 42-50\%~\cite{getafix,bloomberg_apr}. These figures emerged from controlled studies where software engineers were predisposed to reviewing and integrating the suggested changes. Compared to controlled studies, real-world deployments might yield even lower results. For example, an automated refactoring tool that generated blind PRs for feature flag~\cite{featureflagmartinfowler} removal recently deployed at Bloomberg exhibited a considerably lower acceptance rate of 5-8\% in day-to-day development workflows. This result highlights the crucial need to enhance the presentation and relevance of APR suggestions. Software engineers are more likely to decline or overlook patches that appear irrelevant or out of context. Unclear suggestions not only diminish the perceived value of automated patch suggestions as a whole but also amplify the existing burdens that software engineers face, making engineers less likely to continue engaging with APR tools.

With this in mind, this paper describes Bloomberg’s efforts to address the critical problem of optimising the timing, targeting, and relevance of automated patch suggestions. We posit that for a patch to be effective, it must satisfy the following \emph{three relevance criteria}. It must be presented (1) at the right time: presenting a patch during an unrelated task can be distracting, and thus patches should be suggested when software engineers are most receptive, ensuring the least disruption to their workflow; (2) to the right person: not all software engineers are familiar with all parts of a codebase, so patches should be aimed at those with the relevant expertise and domain knowledge; and (3) in the right context: patches should be contextually relevant to the ongoing work and code changes.

To satisfy the above criteria, Bloomberg started a multi-step effort to improve how APR suggestions are presented to software engineers. First, we conceptualised a new tool, B-Assist, that generates patch suggestions within the context of existing PRs which software engineers are already working on (as opposed to generating new blind PRs). We conducted a user study with 34 Bloomberg software engineers to evaluate their opinions regarding this concept, as well as to identify implementation priorities. The study showed that most software engineers prefer receiving suggestions within the context of an existing PR rather than a separate new PR and that "seamless integration through the GitHub pull request user interface" and "having the choice to accept or reject fix suggestions" were key features to prioritise in B-Assist's implementation.

Based on the results from the first study, we implemented and deployed a prototype of B-Assist that uses GitHub's Suggested Changes feature~\cite{githubsuggestedchanges} as the interface to present patch suggestions to software engineers and multiple APR tools as the back-end fix providers. To ensure contextual relevance and timeliness of suggestions, B-Assist only generates suggestions for lines modified in the given PR and in their immediate vicinity. We conducted multiple prototype demonstration sessions with a total of 25 software engineers, followed by a comprehensive user study. This study aimed to identify the specific types of suggestions that software engineers would find most valuable from the tool. The results provided crucial insights, shedding light on the preferences and priorities of software engineers regarding the suggestions they wish to receive.


Following the analysis of the demonstration study results, we iteratively refined our prototype. Subsequently, we conducted a final, interview-based study involving 11 software engineers. This study yielded compelling evidence supporting the efficacy of our approach, showcasing a high acceptance rate of APR suggestions. The suggestions presented by B-Assist were accepted by participants through the GitHub user interface 74.56\% of the time and were given usefulness ratings of at least 4 out of 5 in 78.2\% of cases.

This work makes the following contributions:

\begin{itemize}
\item Introduces a user-centric approach for presenting relevant patch suggestions in a timely manner.
\item Designs, implements, and deploys a prototype of this approach, B-Assist.
\item Presents a multi-stage evaluation of this approach through user studies among Bloomberg software engineers.  
\end{itemize}

This article is structured as follows. In \Cref{sec:bassist_concept}, we introduce the concept of B-Assist. In \Cref{sec:concept_user_study}, we outline the design, analysis, and findings of our user study investigating the concept's viability. Next, in \Cref{sec:bassist_design_deployment}, we introduce our implementation of the B-Assist prototype and its deployment. \Cref{sec:prototype_user_study} details the design, analysis, and outcomes of our prototype user study to confirm B-Assist's potential value in practice. Finally, we close by presenting key takeaways and discussing related work.

\section{B-Assist Concept}
\label{sec:bassist_concept}

\begin{figure}[t]
\includegraphics[scale=0.70]{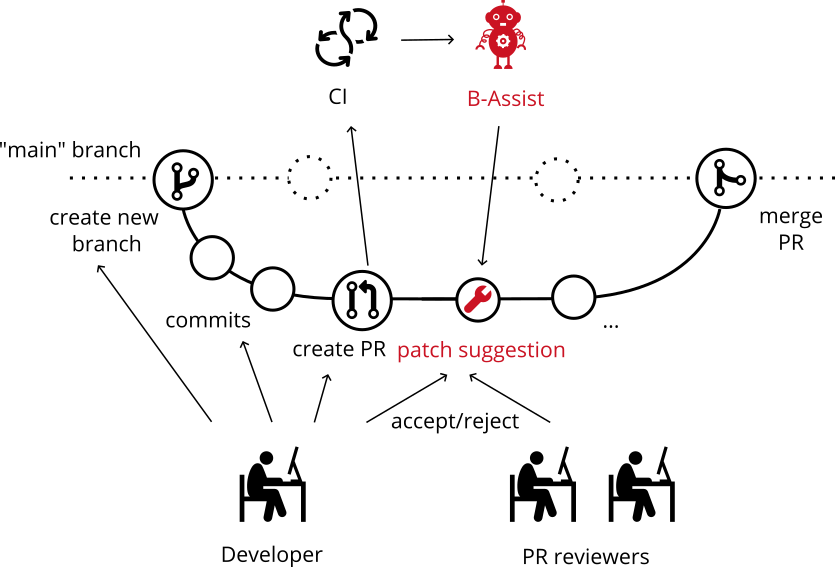}
\caption{B-Assist generates patch suggestions on active pull requests for the lines modified in each commit and their surrounding context.\vspace{-0.1cm}\label{fig:concept}}
\end{figure}

The concept of B-Assist can be traced back to a previous study on developer-centered APR conducted at Bloomberg~\cite{bloomberg_apr}. Within this research, an idea surfaced: the evolution of a program repair tool into a PR reviewer. In other words, an APR tool that could provide suggestions on pull requests opened by software engineers instead of creating its own. One participant in this earlier study remarked that the \textit{``it [PR --- authors' remark] is a very natural place to interact with people, and to raise the problems and also for the relationship between a fix and the problem detection''}. Corroborating this viewpoint, another participant added \textit{``PRs are good [...] because you know that the developer has been working on some parts and you know that if you suggest a fix on this part they will be very likely to engage with the suggestion, so I think that’s a very good time''}.

As illustrated in \Cref{fig:concept}, B-Assist embeds itself into the conventional GitHub flow~\cite{githubflow}. As in the typical GitHub flow employed by many companies, including Bloomberg, when a software engineer adds or modifies a particular program feature, the first step involves creating a new branch. Following their preliminary commits, they create a pull request. This triggers the continuous integration pipeline, which initiates a build and concurrently executes program analysis and repair tools. At this time, other software engineers are also assigned the role of PR reviewers to manually inspect the code changes. Once the CI pipeline is completed, B-Assist activates. It harnesses the data from the CI run, including bug rectifications, format corrections, and the like, generated by CI tools enabled in the target repository. With these tool-generated patches, it checks specifically which of these are targeting code modified by the engineers in the current PR and converts them into suggestion review comments, which can be accepted or rejected as a part of the PR review process. This process is repeated for each subsequent commit in the pull request.

B-Assist meets the three relevance criteria described in \Cref{sec:intro} for the following reasons. First, it presents suggestions at the right time, that being immediately when a software engineer creates a PR and expects to receive feedback on their newly-written code. Second, B-Assist directs patch suggestions to the right audience since the relevant developers and reviewers are already assigned to the PR. Third, B-Assist generates suggestions in the right context, as they are restricted to the code fragments that software engineers are actively working on. The intuition behind this concept is that satisfying the three criteria increases the likelihood that software engineers will review and accept the patch suggestions. In subsequent sections, this belief is backed by empirical evidence.

\section{Soliciting Software Engineer Opinions on B-Assist's Concept}
\label{sec:concept_user_study}

\begin{figure}[tb]
    \centering
    \begin{subfigure}[b]{\columnwidth}
        \centering
        \includegraphics[width=\columnwidth]{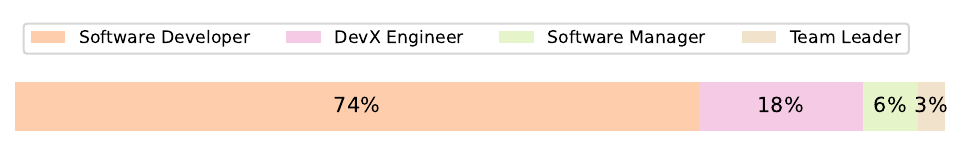}
        \caption{What is your job title? (Pick the closest to your title.).}
        \label{fig:role}
    \end{subfigure}
    \begin{subfigure}[b]{\columnwidth}
        \centering
        \includegraphics[width=\columnwidth]{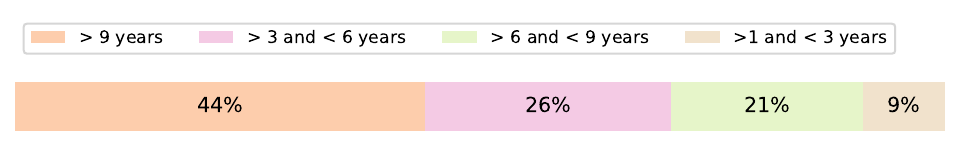}
        \caption{How many years of programming experience do you have?}
        \label{fig:progexp}
    \end{subfigure}
    \begin{subfigure}[b]{\columnwidth}
        \centering
        \includegraphics[width=\columnwidth]{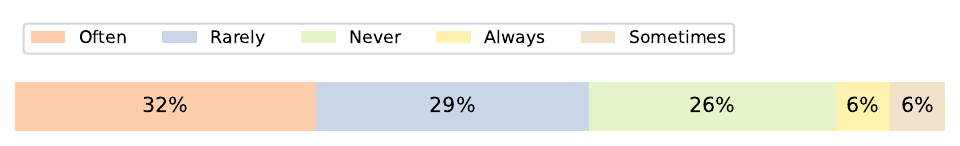}
        \caption{How often do you work with Automated Program Repair (APR) tools? }
        \label{fig:frequency}
    \end{subfigure}
    \begin{subfigure}[b]{\columnwidth}
        \centering
        \includegraphics[width=\columnwidth]{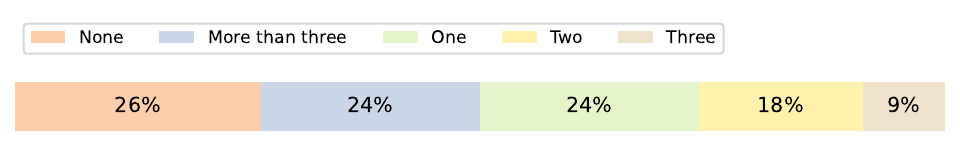}
        \caption{How many APR tools have you worked with?}
        \label{fig:aprtools}
    \end{subfigure}
    \begin{subfigure}[b]{\columnwidth}
        \centering
        \includegraphics[width=\columnwidth]{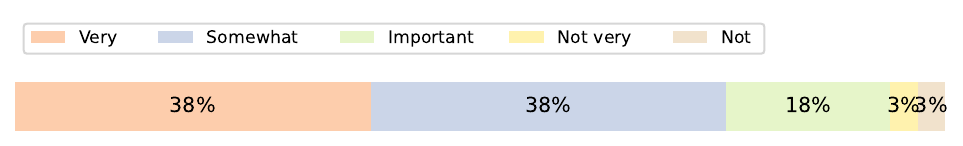}
        \caption{How important do you think automated bug fixing is in terms of speeding up the code review process?}
        \label{fig:importance}
    \end{subfigure}
    \caption{Survey sample background information for the B-Assist concept study, 34 respondents.}
    \label{fig:background_questions}
\end{figure}






\begin{figure}[tb]
    \centering
    \begin{subfigure}[b]{\columnwidth}
        \centering
        \includegraphics[width=\columnwidth]{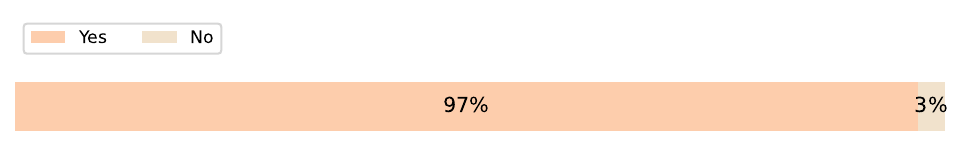}
        \caption{Does this tool seem useful to you?}
        \label{fig:usefulness}
    \end{subfigure}
    \begin{subfigure}[b]{\columnwidth}
        \centering
        \includegraphics[width=\columnwidth]{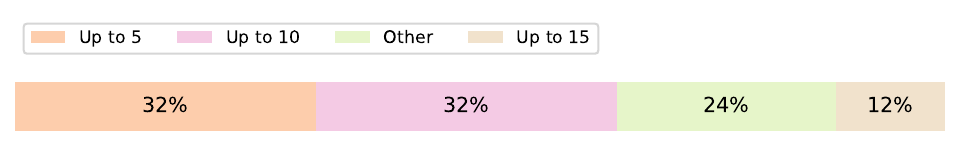}
        \caption{B-Assist aims to provide suggestions only for code that is currently being worked on in a pull request. Given you are reviewing a pull request with roughly 100 lines of modified code, how many suggestions would you be willing to review? }
        \label{fig:suggestions}
    \end{subfigure}
    \begin{subfigure}[b]{\columnwidth}
        \centering
        \includegraphics[width=\columnwidth]{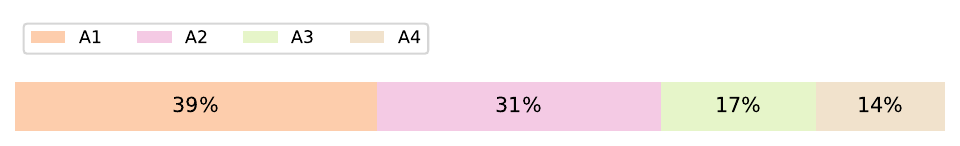}
        \caption{Given the opportunity to apply automatically generated bug fixes directly within Github, what would be your preferred level at which to perform bug fixing during code review? A1: Within the scope of an existing pull request (through GitHub UI). 
        A2: Either would be fine (no particular opinion). 
        A3: Within your IDE. 
        A4: For any part of the project by creating a new pull request (through GitHub UI). 
        }
        \label{fig:visual}
    \end{subfigure}
    \begin{subfigure}[b]{\columnwidth}
        \centering
        \includegraphics[width=\columnwidth]{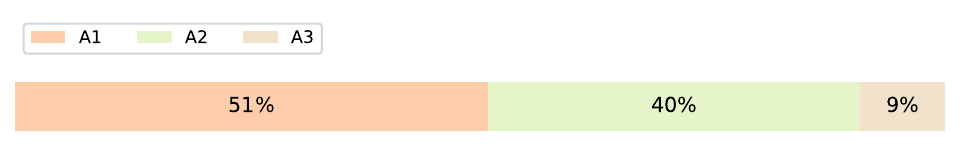}
        \caption{Which features of the tool would most incentivise you to utilise it?
        A1: Seamless integration through the GitHub pull request user interface. 
        A2: Having the choice to accept or reject fix suggestions.                           
        A3: Other. 
        }
        \label{fig:features}
    \end{subfigure}
    \caption{Responses to the survey in the B-Assist concept study, 34 respondents.}
    \label{fig:preliminary_study_feedback}
\end{figure}




We conducted a preliminary user study involving Bloomberg employees to gather initial opinions on the B-Assist concept and understand whether Bloomberg software engineers would be open to adopting such a tool in their workflows.
To control the scope and ensure the data we collected was relevant to our work, we identified three primary goals for this study by considering our concept's proposed functionality and success criteria. The goals of this study were to (1) understand whether a tool like B-Assist would enable a higher patch acceptance rate from engineers, (2) gather initial feedback on how fixes should be presented to software engineers, and (3) identify concrete implementation priorities. This section discusses our approach to designing the user study and the resulting insights we gained.

\subsection{Methodology} \label{sec:preliminary_study_methodology}

To achieve the goals described above, we designed a mixed-method \cite{mixed_method} survey to be distributed to software engineers to refine B-Assist's concept before implementing a prototype of the tool.
The survey consisted of 15 questions, the first eight of which elicited non-sensitive background information about the participants to better understand the survey sample, such as years of engineering experience, typical bug-fixing approaches, and familiarity with APR tools. 
Following these, the survey included a description of B-Assist's early concept\footnote{Description: "Our team is creating a tool, B-Assist, to support software engineers in implementing fixes suggested by APR tools directly through the GitHub pull request UI (webpage). Utilising GitHub’s “Suggested Changes” feature combined with the outputs of APR tools included in the CI/CD pipeline, B-Assist creates review comments providing software engineers with an option to apply automatically generated fixes with the click of a button, adding a new commit to their working branch with their desired fixes."} and asked a number of closed-ended questions to understand how best to present automatically generated patches while taking user experience into account. 
Specifically, these questions aimed to address whether software engineers would find B-assist useful and where they would prefer to fix code issues (e.g., within their IDE, current pull request UI, etc.), addressing goals (1) and (2), respectively. To further tackle goal (2), participants were also asked how many suggested changes they would generally be willing to review. Finally, for goal (3), participants were asked which of the described features would most incentivise them to use B-Assist, providing clear implementation targets.

An electronic version of the survey was distributed internally at Bloomberg, targeting multiple teams from relevant departments (such as Developer Experience, or DevX), who could immediately benefit from using our tool and provide insights based on their prior experience deploying successful developer tools at Bloomberg.

\subsection{Findings} \label{sec:preliminary_study_findings}

A total of 34 participants took part in this survey. Figure~\ref{fig:background_questions} and Figure~\ref{fig:preliminary_study_feedback} summarise the responses to questions about engineer background and on the concept of B-Assist, respectively.

Most respondents (92\%) are software/DevX engineers (Figure \ref{fig:role}), and 91\% of the respondents have at least 3 years of programming experience (44\% of whom notably have over 9 years). Thus, only 9\% have between 1 and 3 years of experience (Figure \ref{fig:progexp}).
From Figure \ref{fig:frequency}, we observe that 55\% of the respondents have rarely (29\%) or never (26\%) worked with APR tools, whereas the remaining 45\% is divided between those who work with APR tools often (32\%), sometimes (6\%), or always (6\%). It should be noted that, in this study, APR tools were defined as any tools that generate patch suggestions, which includes both research APR tools~\cite{goues2019automated} and popular industrial tools, such as clang-tidy~\cite{clangtidy}, that generate patch suggestions for identified static analysis violations. Regardless, we can see from our sample that the engineers' experience with APR tools is not widespread, as 24\% of respondents declared they have used only one tool, 18\% two tools, 9\% three tools, and 24\% more than three tools (Figure \ref{fig:aprtools}). 
Among those who have experience with APR, the most frequently used tools are clang-tidy, clang-format~\cite{clangformat}, and in-house clang tools developed using Bloomberg's clang-metatool framework~\cite{clangmetatool}, as well as other linters such as gcc-lint.
Still, 94\% of participants generally express positive feelings towards the importance of automated bug fixing in terms of speeding up the code review process (Figure \ref{fig:importance}), with only a small minority of participants deeming it not very useful (3\%) or not useful at all (3\%). 

When presented with a general description of the B-Assist concept, the vast majority of respondents (97\%) felt this tool would be useful to them. 
Regarding how many automatically-generated suggestions they would like to receive while reviewing a pull request modifying about 100 lines of code, the participants were equally split with ``up to 5" (32\%) and ``up to 10" (32\%), while 12\% opted for a higher upper bound of 15 suggestions and 24\% had other preferences, the most frequent of which was to include all suggestions, provided they are relevant.

\begin{framed}
	\vspace{-5px}
\begin{finding}\label{finding:number_of_suggestions}
Almost all respondents felt that the tool would be useful in their workflows, but the majority would prefer reviewing relatively few suggestions (up to 5 or 10).
\end{finding}
	\vspace{-5px}
\end{framed}

Moreover, regarding the participants' feedback on the possibility of applying automated patches directly within GitHub, only 12\% indicated that they would be happy to see the patch applied to any part of the project by creating a new PR. In contrast, the largest minority stated they would like to see them applied at a finer-grain level closer to their work, i.e., only within the scope of an existing PR (36\%). A significant group (31\%) stated no preference, and the remainder preferred handling these patches only within their IDE (17\%). Finally, when asked about the most desirable feature to them among those proposed in the B-Assist concept (i.e., the features which would most incentivise them to use B-Assist), 51\% of participants expressed a preference for seamless integration through the GitHub PR user interface, and 40\% valued the ability to accept/reject patch suggestions. 

\begin{framed}
	\vspace{-5px}
\begin{finding}\label{finding:impementation_priorities}
The majority of participants preferred receiving suggestions within the context of an existing PR rather than opening a new PR, and the seamless integration in the PR user interface combined with the ability to easily accept/reject the suggestions were repeatedly stated as the most desirable features.
\end{finding}
	\vspace{-5px}
\end{framed}

\section{B-Assist: User-Centric Design and Deployment}
\label{sec:bassist_design_deployment}

The results of the preliminary user study presented in \Cref{sec:concept_user_study} motivated the design and implementation of B-Assist. During the design phase, a particular emphasis was placed on how relevant patches should be seamlessly presented to software engineers through the GitHub pull request user interface.

\subsection{User Experience}

\begin{figure}[t]
    \centering
    \includegraphics[width=1\linewidth]{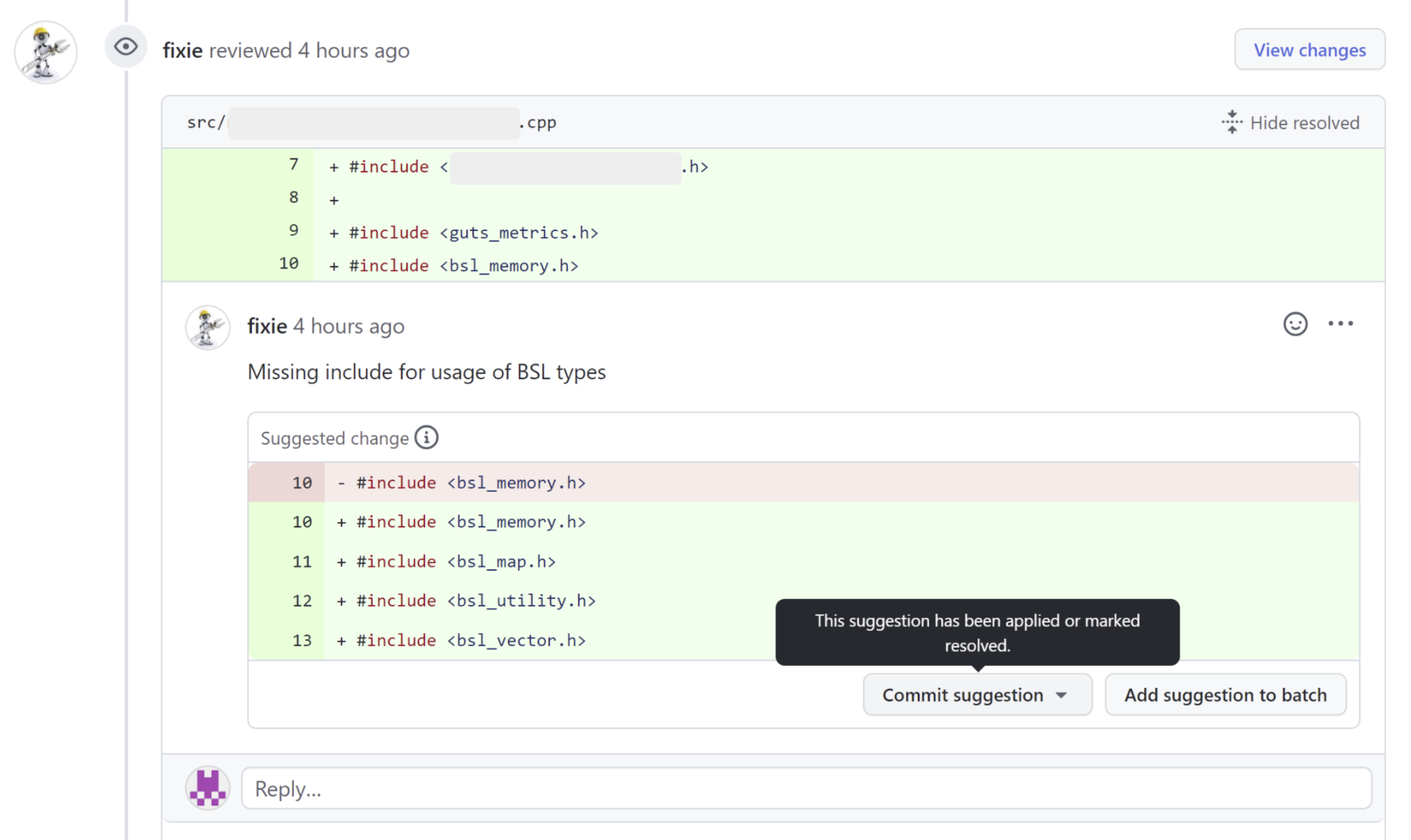}
    \caption{A real-world patch suggestion generated by B-Assist that was accepted by a software engineer at Bloomberg.}
    \label{fig:accepted_b-assist_suggestion}
\end{figure}

As the core user interface element for B-Assist to present patches to software engineers, we adopted GitHub's Suggested Changes~\cite{githubsuggestedchanges}, a relatively new feature that is already used effectively in industry by software engineers and team leaders to provide feedback on pull requests. Prior work has also explored using this feature in combination with code review bots~\cite{palvannan2023suggestion, chbrown_suggested_changes}. As illustrated in \Cref{fig:accepted_b-assist_suggestion}, a suggested change is displayed in the context of a given PR's discussion, and software engineers can accept or reject the suggestions with a single click. If a change is accepted, a new commit containing the change is automatically appended to the pull request. If it is rejected, it is hidden from the discussion. Thus, this interface meets the software engineers' priorities according to \Cref{finding:impementation_priorities}.

To differentiate repairs from their presentation, we use the following terminology. A \emph{fix} or a \emph{patch} is the source code modification to address an issue or repair a bug. A \emph{patch suggestion} is the representation of a patch provided by B-Assist to software engineers via the GitHub Suggested Changes user interface. 

\subsection{Implementation}
\label{sec:limitations}
\begin{figure}
    \centering
\includegraphics[width=0.55\linewidth]{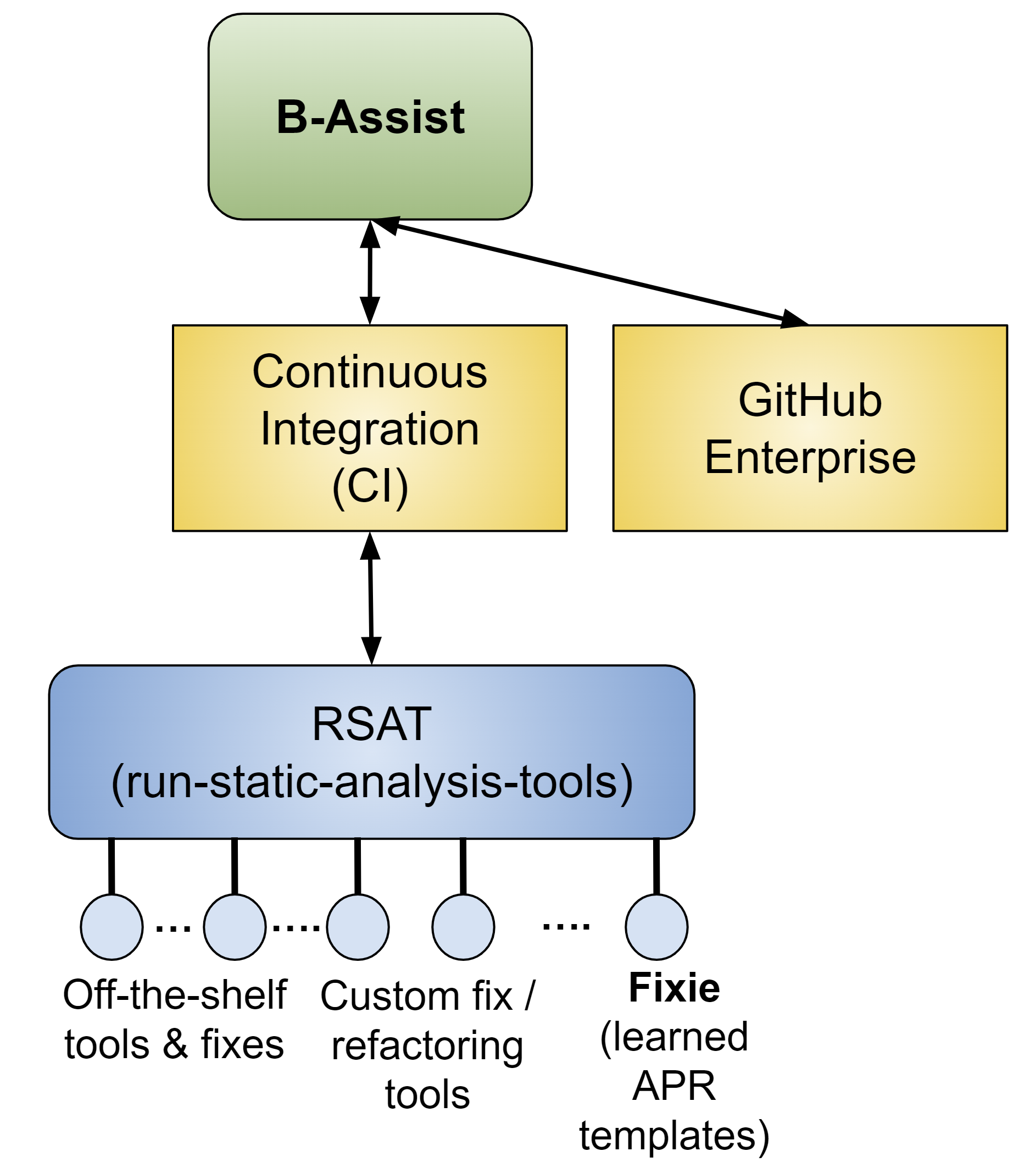}
    \caption{B-Assist connects two main components: Bloomberg's continuous integration service that runs various APR tools via the RSAT wrapper, and GitHub, which is used to present patch suggestions to software engineers.}
    \label{fig:architecture}
\end{figure}

The architecture of the solution can be seen in Figure \ref{fig:architecture}. B-Assist integrates with Bloomberg's continuous integration (CI) service and GitHub Enterprise (Bloomberg's version control and code review platform). When software engineers create a pull request or modify an existing one, a CI build is automatically triggered. Next, B-Assist receives the results from Run Static Analysis Tools (RSAT), an umbrella tool enabled in the CI pipeline that provides a unified entry point to multiple static analysis and APR tools used at Bloomberg. RSAT creates centralised reports and patches containing the output generated by various program analysis and repair tools, including Fixie, a template-based APR tool that learns fix patterns from software engineers' fixes~\cite{fixie, bloomberg_apr}. 

B-Assist is implemented as a GitHub app, a type of integration that can extend GitHub functionality by leveraging GitHub's REST API~\cite{githubrest}. B-Assist listens for when CI checks are completed and identifies their status and output. Using the output of the RSAT check, it accesses the corresponding report for the given PR to check if relevant patches were produced during the CI run. If so, B-Assist processes these new patches and uses the REST API to post review comments containing patch suggestions to the PR.

RSAT produces patches in the Unidiff format~\cite{unidiff} used by various utilities such as \texttt{patch}, \texttt{diff} and \texttt{git}. Meanwhile, GitHub Suggested Changes uses an ad-hoc format that only allows the replacement of several consecutive lines within a Unidiff representation of pull request commits. Thus, B-Assist implements a converter that takes in (1) a Unidiff representation of a pull request commit and (2) a Unidiff representation of the generated patch, and outputs the patch in the Suggested Changes format.

B-Assist is programming language agnostic. Thus, it can interact with any CI process running on GitHub PRs and any APR tool if it produces patches in the Unidiff format. The initial release of B-Assist focused specifically on C++. This choice was driven by the language's pivotal role in the development of software solutions at Bloomberg~\cite{cplusplusBloomberg}. Bloomberg has tens of thousands of C++ projects and thousands of C++ libraries to reflect its use cases.

Some noteworthy limitations have been identified in our current implementation. Due to the current technical limitations of the Suggested Changes format, not all patches can be conveniently displayed through this interface. For example, the Suggested Changes API only permits inserting a change at a single location. Although it allows for modifying multiple consecutive lines, a change affecting multiple non-adjacent lines cannot be displayed as a single suggestion. We expect this limitation will be addressed in future versions of GitHub. Another limitation of the current implementation is that it does not account for the difference in the behaviour of different APR tools. For example, some tools split a single repair into multiple separate patches if the distance between individual changes is large. In this case, B-Assist may only present one and ignore the other parts of the repair. Extensions to B-Assist providing tighter integration with specific APR tools could be implemented in future to address this limitation.

\subsection{B-Assist's Journey to Deployment}
The B-Assist project spanned nine months, starting from its conceptualisation to the launch of the initial release on October 1, 2023. B-Assist is the result of a collaborative effort between industry and academia, involving a partnership between Bloomberg and University College London (UCL). During this time, both entities collaborated closely, contributing expertise and resources throughout all stages of its development. UCL members were onboarded as Bloomberg software engineering researchers to get hands-on experience and conduct user studies. After defining the B-Assist concept (Section \ref{sec:bassist_concept}), we solicited opinions from potential users at Bloomberg and refined it based on their feedback (Section \ref{sec:bassist_design_deployment}).
A prototype was subsequently built and deployed to one GitHub organisation containing 40 C++ repositories owned by one team. Next, we solicited feedback on the prototype (Section \ref{sec:prototype_user_study}) and its suggested patches (Section \ref{sec:concluding_study_interview}). Given the positive responses, we worked towards deploying B-Assist more widely across other GitHub organisations within Bloomberg. 
As of October 1, 2023, B-Assist was made available to all software engineering teams, meaning they can install the B-Assist GitHub App to their organisations or repositories. Software engineers who would like to try B-Assist can follow a few steps documented in the central Bloomberg Engineering documentation. 
In the first week after deployment, eight teams had already adopted B-Assist and successfully accepted several suggestions, an example of which can be seen in Figure \ref{fig:accepted_b-assist_suggestion}.

\section{Soliciting Software Engineers' Opinion on B-Assist Prototype}
\label{sec:prototype_user_study}
Once a functional prototype of B-Assist, according to the concept described in Section \ref{sec:bassist_design_deployment}, was ready for deployment, we deemed further investigation involving software engineers was necessary to (1) guarantee that B-Assist stands to provide a significant positive impact on the software development process at Bloomberg and (2) learn more about how software engineers want to interact with B-Assist. This was achieved by conducting multiple tool demonstration sessions followed by a survey distributed to participating Bloomberg software engineers. The following subsections describe the methodology and findings of this study.

\subsection{Methodology} \label{sec:concluding_study_demo_methodology}
To encourage the uptake of B-Assist and gather more concrete feedback, we invited several software engineering teams at Bloomberg to participate in tool demonstration sessions. These consisted of a slideshow presentation explaining how B-Assist could reduce the bug-fixing effort for Bloomberg engineers, followed by a live demonstration of the tool in a test repository. Once the participants were given a chance to experience the functionality of B-Assist and view the proposed presentation of patch suggestions, a follow-up survey comprised of three qualitative (closed-ended) questions and an open-ended question was distributed. This survey included questions covering various key topics on which software engineers could provide specific feedback, such as fix presentation, additional features or considerations, and what types of code issues they would like the tool to address (e.g., code formatting, missing/unused imports, potential null pointer references, etc.). The final open-ended survey question encouraged participants to make any additional comments they felt would be relevant.
The participants in this study were gathered through convenience sampling by advertising this study within Bloomberg Engineering teams who are primarily developing using C++. This decision was made because the B-Assist prototype was initially deployed to 40 C++ Bloomberg repositories.

\subsection{Findings}
\label{sec:concluding_study_demo_feedback}
A total of 25 participants took part in the demo sessions and answered the survey: 17 Software Engineers, six DevX Engineers, and two Engineering Managers. Their expertise varies as follows: over 9 years (11 participants), 6-9 years (6 participants), 3-6 years (7 participants), and
1-3 years (1 participant).
The first survey question asked \textbf{what types of issues they would like B-Assist to present patch suggestions for}. 
The majority of participants pointed out missing/unused imports, null dereferences, and formatting issues (in this order).
However, participants stressed that patch suggestions regarding formatting issues would only be desirable if they point out issues not fixed by existing auto linters (e.g., linterator) and are
applicable in bulk rather than one at a time. In the words of one participant who captured others' concerns, "\textit{formatting would be interesting, although applying formatting changes one at a time sounds rather painful}".
Some less frequent (yet noteworthy) observations were made about providing B-Assist with an option allowing software engineers to select their preferred suggestion types and limit patch suggestions to those only generated by specific RSAT driver(s) a team is interested in using. Others expressed that they wanted to have the ability to filter B-Assist patch suggestions based on fix severity.
Several software engineers were willing to receive patch suggestions for as many types of issues as possible (e.g., all those that can be found by a static analyser) as long as the issues were introduced in (and therefore relevant to) the current PR.
Finally, some suggested including patch suggestions for common programming language pitfalls/deprecated API usage (e.g., "\textit{certain types of issues in Python, such as default list arguments or in shell}").

\begin{framed}
	\vspace{-5px}
\begin{finding}\label{finding:issue_types}
    The participants listed missing/unused imports, null dereferences, and formatting issues as the types of issues which they most want to receive automated suggestions for.
  \end{finding}
	\vspace{-5px}
\end{framed}

The second question asked software engineers \textbf{what additional features or changes they would recommend to improve the B-Assist user experience}.
The majority of the responses fell under two main requests: indicating the reason for the change and including which APR tool generated the suggestion.
This feedback highlights that, for each patch suggestion, software engineers would like to have a clear reason why the fix is being proposed (or at least \textit{add an indication whether it is an error or a warning and whether the fix is required or optional}). They would also like an indication as to which RSAT driver generated the fix as this might, for example, \textit{"help people replicate locally if they wanted (e.g., by running run-static-analysis-tools --tool xyz)"}. These findings are consistent with previous studies~\cite{noller2022trust}, which identified that providing additional information, such as patch explanations, enhances software engineers' trust in APR.

\begin{framed}
	\vspace{-5px}
\begin{finding}\label{finding:reason}
     The additional features most desired by participants for the patch suggestion generator are (1) to provide the reason for the change, and (2) specifying its source, i.e., which APR tool generated the fix.
  \end{finding}
	\vspace{-5px}
\end{framed}

Based on these findings, we incorporated both features into our prototype. However, the RSAT tools available at Bloomberg can only provide limited explanations for the generated patches, so additional future work is needed in this respect.

The last question asked software engineers for additional comments regarding the tool, if they had any. Many enthusiastic and positive comments were left (e.g., \textit{``We've been needing something like this for a while. Nice work!''}, \textit{``Looks really cool!''}, \textit{``Super cool, let's figure out a transition plan so it can get used around Bloomberg.''}) and a few technical suggestions were also shared, such as \textit{``I think it is a good idea and might benefit from a config to control what is happening on a repo. It would be cool to also have a way to opt in as a user when committing to repos that don't have it enabled for the scenario where I issue a bulk PR to 100s of repos.''} and \textit{ ``I am worried that filtering out lines that are not changed might produce incomplete fixes.''} \Cref{sec:discussion} discusses these and other limitations, as well as possible improvements.

\section{Soliciting Software Engineers' Opinion on B-Assist Patch Suggestion}
\label{sec:concluding_study_interview}

\begin{table*}[t]
    \centering
    \caption{Summary of results of the study on B-Assist patch suggestions. For each patch suggestion, we report (1) the source of the issue and the type of the fix RSAT provided, (2) the percentage of participant answers for how they would respond to the patch (from the four possible actions), (3) the percentages of responses representing the usefulness of the patch  (on a Likert scale from 1 to 5), and (4) the percentage of participants who prefer a patch suggestion against those who prefer a text warning. 74.56\% (65.46\% + 9.1\%) of suggestions were accepted from within the GitHub UI either directly or after a modification, and participants rated suggestion usefulness at least 4 out of 5 78.2\% of the time. Finally, while there are differing opinions on some suggestions, participants prefer being presented with a patch suggestion over a textual warning 89\% of the time.}
    \label{tab:combined}
    \resizebox{\textwidth}{!}{
    \begin{filecontents*}{data/results_with_average_and_format.csv}
Suggested Fix,Type,Source,Accept,Modify and Accept,Fix my own way,Ignore/Reject,5 (Very Useful),4,3,2,1 (Not Useful),Code Suggestion,Text Warning
No.1,Modernising Language Standards,Pre-existing Issue,90.9,0,9.1,0,63.6,27.3,0,9.1,0,90,10
No.2,Changing Function Definition,Fabricated Snippet,18.2,18.2,63.6,0,36.4,18.2,36.4,9.1,0,70,30
No.3,Performance Improvement,Pre-existing Issue,27.3,27.3,0,45.5,27.3,18.2,18.2,9.1,27.3,90,10
No.4,Removing Dead Code,Pre-existing Issue,18.2,9.1,72.7,0,54.5,45.5,0,0,0,90,10
No.5,Best Practice,Fabricated Snippet,27.3,36.4,36.4,0,54.5,9.1,36.4,0,0,70,30
No.6,Readability,Fabricated Snippet,72.7,0,18.2,9.1,45.5,27.3,18.2,9.1,0,80,20
No.7,Formatting,Introduced in Existing Code,100,0,0,0,54.5,18.2,27.3,0,0,100,0
No.8,Formatting,Introduced in Existing Code,100,0,0,0,54.5,27.3,18.2,0,0,100,0
No.9,Dependency Management,Introduced in Existing Code,100,0,0,0,81.8,18.2,0,0,0,100,0
No.10,Dependency Management,Introduced in Existing Code,100,0,0,0,90.9,9.1,0,0,0,100,0
Average,,,65.46,9.1,20,5.46,56.35,21.84,15.47,3.64,2.73,89,11
\end{filecontents*}
\pgfplotstabletypeset[
    col sep=comma,
    column type=r,
    string type,
    columns={Suggested Fix, Source, Type, Accept, Modify and Accept, Fix my own way, Ignore/Reject,5 (Very Useful),4,3,2,1 (Not Useful), Code Suggestion, Text Warning
    },
    columns/Suggested Fix/.style={column name=Suggested, column type={|l|}},
    columns/Type/.style={column name={Fix Type}, column type={|l}},
    columns/Source/.style={column name={Issue Source}, column type={l}},
    columns/Accept/.style={column type={||r|}},    
    columns/Modify and Accept/.style={column name=Modify},
    columns/Fix my own way/.style={column name=Fix my},
    columns/Ignore/Reject/.style={column name=Ignore/},
    columns/5 (Very Useful)/.style={column name=5, column type={|r|}},
    columns/1 (Not Useful)/.style={column name=1, column type={r||}},
    columns/Code Suggestion/.style={column name=Patch},
    columns/Text Warning/.style={column name=Warning, column type={r|}},
    every head row/.style={
    before row=\hline \multicolumn{3}{|c||}{} & \multicolumn{4}{c||}{Acceptance} &\multicolumn{5}{c||}{Usefulness} &\multicolumn{2}{c|}{Preferred Format} \\ \hline ,
    after row={
    Fix & & & & \& Accept & own way & Reject & Very& & & & Not &  &  \\
    & &     & & &  & & Useful& & & & Useful & & \\
    \hline
    },
    },
    every row 0 column 3/.style={ postproc cell content/.style={ @cell content/.add={$\bf}{$} } },
    every row 1 column 5/.style={ postproc cell content/.style={ @cell content/.add={$\bf}{$} } },
    every row 2 column 6/.style={ postproc cell content/.style={ @cell content/.add={$\bf}{$} } },
    every row 3 column 5/.style={ postproc cell content/.style={ @cell content/.add={$\bf}{$} } },
    every row 4 column 5/.style={ postproc cell content/.style={ @cell content/.add={$\bf}{$} } },
    every row 5 column 3/.style={ postproc cell content/.style={ @cell content/.add={$\bf}{$} } },
    every row 6 column 3/.style={ postproc cell content/.style={ @cell content/.add={$\bf}{$} } },
    every row 7 column 3/.style={ postproc cell content/.style={ @cell content/.add={$\bf}{$} } },
    every row 8 column 3/.style={ postproc cell content/.style={ @cell content/.add={$\bf}{$} } },
    every row 9 column 3/.style={ postproc cell content/.style={ @cell content/.add={$\bf}{$} } },
    every row 10 column 3/.style={ postproc cell content/.style={ @cell content/.add={$\bf}{$} } },
    every last row/.style={before row=\hline, after row=\hline},
    every column/.style={column type/.add={}{|}},
    ]
{data/results_with_average_and_format.csv}

    }
\end{table*}

While the initial surveys from the preliminary study (Section \ref{sec:concept_user_study}) and tool demonstration sessions (Section \ref{sec:prototype_user_study}) were successful in identifying and confirming engineer interest in B-Assist's concept, it was also crucial to investigate B-Assist's potential use cases more thoroughly. For instance, what types of fixes should be presented by B-Assist and how engineers would behave in different scenarios when reviewing suggested change comments on their pull requests are both important considerations which need to be understood to ensure that B-Assist can be utilised to its full potential. To this end, we designed a user study where participants were shown examples of patch suggestion comments and asked a series of questions regarding how they would react.

\subsection{Methodology}
\label{sec:concluding_study_interview_design}
To design this study, we first selected a set of patch suggestions generated by B-Assist to be presented to the participants. Informed by the findings from the demonstration sessions (Section \ref{sec:concluding_study_demo_feedback}), and based on the types of patches provided by the APR tools available to B-Assist through RSAT, we selected ten examples of fixes representing a wide range of issues that B-Assist could help rectify. These types are listed in the column ``Fix Type'' in \Cref{tab:combined}. They are selected from broader categories of issues identified as the most asked for by Bloomberg software engineers, as explained in \Cref{finding:issue_types}.

Code fragments involving issues corresponding to the above types were selected from various sources. Three were taken directly from the repositories managed by co-authors of this study at Bloomberg (No. 1, No. 3 and No. 4 in \Cref{tab:combined}). These repositories did not include issues of other types, and searching for such issues in repositories managed by other teams at Bloomberg was impractical since that would require not only getting access to the source code in these repositories but also modifying their CI configuration, which might have significantly disrupted their development workflows. Thus, we created four buggy fragments by intentionally introducing issues in functional code (No. 7, No. 8, No. 9 and No. 10 in \Cref{tab:combined}). The final three (No. 2, No. 5 and No. 6 in \Cref{tab:combined}) were written from scratch by using the documentation of the analysis and repair tools providing the fix as a representative usage example.

We then pushed these code fragments to a pull request in a mock repository with both B-Assist and RSAT enabled and prompted B-Assist to generate patch suggestions. This process was carried out by one of the authors, and another author from Bloomberg verified all snippets and resulting fixes to ensure they covered a satisfactory range of issues before proceeding. The fixes were then compiled in a survey presented to each participant during a one-to-one interview.

The questionnaire contains two quantitative multiple-choice questions for each fix on (1) how they would respond to the suggested change if it appeared on their pull request and (2) how useful they found the suggestion (on a 5-point Likert scale). Alongside this, as we expect bug-fixing approaches and preferences to vary for each engineer (as indicated by the results discussed in Section \ref{sec:preliminary_study_findings}), we aimed to capture their interpretations more precisely. Thus, after completing the survey, participants were prompted to expand on why they chose a given option in the survey, and the interviewer took notes based on each participant's qualitative feedback. For each fix, the interviewer specifically asked whether they would prefer a suggested change attempting to fix the problem or a simple text comment bringing the issue to their attention. After going through the survey, the interviewer also posed some general questions, gathering participant opinions on fixing bugs in the GitHub user interface, their typical bug-fixing approaches, and the value they believe APR tools stand to provide to the code review process.

\subsection{Findings}
A total of 11 participants (P1--P11) took part in the interviews. Their role, experience and overall reaction to the B-Assist suggestions are summarised in~Table \ref{table:interviews}. \begin{table}[tb]
\caption{Interviews: Participant Role, Expertise, and Reaction to B-Assist in terms of acceptance and usefulness of the suggested patches. For each participant, we summarise their reaction as the average of positive and negative feedback across the 10 suggested patches, where positive feedback comprises \textit{Accept, Modify \& Accept, Fix Own Way} reactions, and \textit{Very Useful, Useful and Somewhat Useful} for acceptance and usefulness, respectively.}
\label{table:interviews}
\resizebox{.45\textwidth}{!}{    
\begin{tabular}{llrrrrr}
\hline	\textbf{ID}	&	\textbf{Role}	& \textbf{Expertise}	&
        \multicolumn{2}{c}{\bfseries Acceptance}	&
         \multicolumn{2}{c}{\bfseries Usefulness}	\\	
\multicolumn{2}{c}{} &		
\textit{(in years)} &
\textit{Pos.}	&	
\textit{Neg.}	&	
\textit{Pos.}	&	
\textit{Neg}\\	
\hline
P1	&	Sw Eng.	&	8	&	80\%	&	20\%	&	80\%	&	20\%	\\	
P2	&	Sw Eng.	&	6	&	90\%	&	10\%	&	70\%	&	30\%	\\	
P3	&	Sw Eng.	&	2	&	100\%	&	0\%	&	100\%	&	0\%	\\	
P4	&	Sw Eng.	&	3	&	100\%	&	0\%	&	100\%	&	0\%	\\	
P5	&	Sw Eng.	&	22	&	100\%	&	0\%	&	100\%	&	0\%	\\	
P6	&	Sw Eng.	&	20	&	100\%	&	0\%	&	100\%	&	0\%	\\	
P7	&	Sw Eng.	&	42	&	100\%	&	0\%	&	100\%	&	0\%	\\	
P8	&	Sw Eng.	&	2	&	100\%	&	0\%	&	100\%	&	0\%	\\	
P9	&	Sw Eng.	&	20	&	90\%	&  10\%	&	90\%	&	10\%	\\	
P10	&	Sw Eng.	&	5	&	90\%	&	10\%	&	100\%	&	0\%	\\	
P11	&	Team Lead	&	26	&	90\%	&	10\%	&	90\%	&	10\%	\\	\hline
\end{tabular}
}
\end{table}
 Ten participants are Software Engineers with experience ranging from 2 to 42 years (mean=12 years, median=7 years), and one is a Team Leader with 26 years of experience. 
Regardless of their expertise, each individual's reaction to B-Assist's suggestions was very positive with respect to both acceptance rate and usefulness, with 80\% to 100\% and 70\% to 100\% positive reactions reported on average, respectively. 
Table~\ref{tab:combined} summarises the participants' feedback for each suggested patch. Half of the fixes suggested by B-Assist were accepted by the vast majority of participants. Notably, 100\% of the participants answered \textit{Accept} for fixes Nos. 7, 8, 9, 10 and 91\% answered \textit{Accept} for fix No. 1. 74.56\% of suggestions were accepted from within the GitHub user interface either directly or after modification. Software engineers had no issue immediately accepting fixes related to code formatting (No. 7 \& 8) and dependency management (No. 9 \& 10). However, several participants did note that they would only accept all four of these fixes \textit{'assuming the CI checks agree with the IDE configurations'} (P5), stressing the importance of configurability, an observation consistent with previous findings when integrating APR tools at Bloomberg~\cite{fixie}. As for the remaining five fixes, two (namely fixes Nos. 5 and 6) received mainly positive reactions (i.e., \textit{Accept} or \textit{Modify \& Accept}).
For fixes Nos. 2 and 4, the majority of participants (66\% and 73\%, respectively) indicated that they would prefer to \textit{fix the code their own way} and the remaining ones would instead \textit{accept} (18\% for fix No. 2, 27\% for fix No. 4) or \textit{modify and accept} (18\% for fix No. 2, 27\% for fix No. 4) the suggested patches. It should be noted that when B-Assist brings an issue to the software engineer's attention, even if they choose to address it in their own way, the software has partially fulfilled its role by making them aware of potential problems or areas of improvement in their code. Participants' feedback supports this, as P7 stated that even \textit{``if he doesn’t accept the guess, it’s good to have the option because he believes that 9/10 times it’s going to be correct and easier to apply it that way''}. Similarly, P9 remarked that \textit{``it’s good as a suggested fix: in general if there are multiple ways of dealing with something, he still likes the suggested fixes because there’s a chance it’s correct. If it’s not correct, no big deal, he can fix it in his own way.''}

In only one case (i.e., fix No. 3), some respondents (45\%) indicated they would reject the suggestion. However, the majority (55\%) indicated that they would still \textit{accept} or \textit{modify and accept it}. For this particular fix, P1 commented that \textit{'this is a test file, so the proposed change doesn’t matter as much. I am unsure why this suggestion is being made/what benefit it provides.'}

\begin{framed}
	\vspace{-5px}
\begin{finding}\label{finding:acceptance_fix_stats}
The suggestions presented by B-Assist were accepted by participants through the GitHub user interface either immediately or after modification 74.56\% of the time. In only 5.46\% of cases the suggestions were completely dismissed, suggesting that instances of irrelevant suggestions were rare.
\end{finding}
	\vspace{-5px}
\end{framed}

\begin{framed}
	\vspace{-5px}
\begin{finding}\label{finding:favourite_fixes}
APR fixes generated for code formatting and dependency management issues were the easiest for engineers to accept when presented as suggestions.
\end{finding}
	\vspace{-5px}
\end{framed}

The participants also rated the usefulness of patch suggestions generated by B-Assist. On average, the majority of participants found the suggestions \textit{very useful} (56.35\%) and \textit{useful} (21.84\%) across the ten fixes. The suggestions had an average usefulness rating of 4.26 out of 5. Fix No. 3 is the only one for which the patch suggested by B-Assist was deemed \textit{not useful} by some participants (27\%).

\begin{framed}
	\vspace{-5px}
\begin{finding}\label{finding:usefulness_fix_stats}
The patch suggestions provided by B-Assist had an average usefulness rating of 4.26 out of 5, receiving a rating of least 4 out of 5 in 78.2\% of cases.
\end{finding}
	\vspace{-5px}
\end{framed}

Although B-Assist suggestions were perceived positively, some participants expressed interesting concerns about the ease with which these tools may allow software engineers to modify the codebase, suggesting a potential for over-reliance or misuse. P2 mentioned that \emph{``the audience is very important and the reviewer requires significant domain knowledge; the audience is as important as the tool; whether it should present fixes depends on the issue and how confident the tool is in its suggestion''}. This comment introduces a new challenge of using APR, suggesting that APR suggestions should be nuanced, contingent on the nature of the given issue and the confidence level of the tool's proposed solutions.

From the interviews, we discovered that in those cases where software engineers would prefer to make a fix their own way or rather ignore the suggested fix, the main reason was the clarity of the suggestion rather than its relevance to their current work. Similar observations about the clarity of automatically generated patches were made in previous user studies on program repair~\cite{noller2022trust}.

In those cases where the participants were feeling extremely positive (fixes Nos. 7--10), the interview notes revealed that they liked the fact that the suggestion reduced the need to context-switch -- e.g. \textit{``Great as a code suggestion to click and accept, would be annoying to have to return to IDE to fix this.''} (P10, fix No. 9), \textit{``This is useful as otherwise the software engineer would need to access the RSAT report manually and go to fix this issue locally.''} (P10, fix No. 10) -- this being precisely one of the primary aims for the proposal of a tool such as B-Assist in Bloomberg. This sentiment is also reinforced by their answers to the non-fix-specific interview questions, of which we report a handful of quotes below:

\begin{itemize}

\item P2: \textit{'I like the way fixes are suggested on pull requests due to the convenience of not having to return to IDE.'}

\item P10: \textit{'It’s good to receive fixes for a particular code while you’re working on it.'} 

\item P2: \textit{'It is helpful, as it acts like an “extra pair of eyes” to ensure software engineers are adhering to best practices.'} 

\item P1: \textit{'If it’s a minor issue, a direct fix in the pull request is great. If more extensive fixes, text-based suggestions warning the software engineer would be more useful. Formatting and missing headers can pretty much always be accepted and are therefore useful as fixes.'} 

\item P8: \textit{'Since the introduction of batching suggestions together, I like it more. It would be a pain to add each change separately in different commits'.} 
\end{itemize}


The feedback gathered during the interviews highlights that software engineers love to "peek under the hood and see the engine". They want to examine the inner workings and details of the generated fixes. For instance, participant P4 suggested 
\textit{'It would be good if more info is provided as part of the fix suggestion. Perhaps even a URL to the specific issue/fix as a reference'}, regarding the usefulness of fix No. 1. Another, P1, stated \textit{'It depends on how the missing dependency is determined. This suggestion is much more useful if it is based on an actual dependency search.'} when discussing fix No. 10, where one of the RSAT drivers deemed it necessary to add a missing dependency. These reactions further motivate the need for more work on patch explainability~\cite{ssbse23ExplainableAPR} to ensure that engineers can immediately understand what is being changed and why.

When conducting the interviews, we asked software engineers if they prefer receiving the output of CI runs as patch suggestions or text warnings describing the problem and its locations in the code. One participant (P11) did not provide information about whether they prefer text warnings or patches. The results of this study are presented in the columns ``Preferred Format'' of Table~\ref{tab:combined}. Overall, the software engineers preferred patch suggestions in 89\% of cases.

\begin{framed}
	\vspace{-5px}
\begin{finding}\label{finding:prefer_fixes}
The majority of software engineers prefer reviewing patch suggestions rather than text warnings pointing to locations in their PR.
\end{finding}
	\vspace{-5px}
\end{framed}

\section{Threats to Validity}
\label{sec:threats}
Throughout this study, we have tried as much as possible to minimise factors which could threaten the validity of our study.
To mitigate the \textit{construct validity} threats in our qualitative study, we carefully designed the questionnaire and interview questions to ensure a unified understanding between what the researcher has in mind and what the respondent eventually understands. While two authors (one from industry and one from academia) led the questionnaires and interview design, these were revised and validated by three other authors: individually at first and then through a collaborative session where consensus was reached. Moreover, a read-aloud pilot study carried out by those authors not involved in the design of the questionnaires ensured the detection and elimination of incompatible terminology and other misunderstandings. The \textit{internal validity} of qualitative studies can be undermined when causal factors are examined and reported. As this study is mainly data-driven with first- and second-degree collection methods (the interviews and questionnaire), we present the results as observed. The open-ended questions were analysed and coded using deductive thematic analysis~\cite{braun2006using} by one author and validated by another in multiple collaborative sessions until a consensus was reached. Similarly, while one author led the interviews, three other authors revised and validated the results to ensure an appropriate analysis of the interview notes. 
Some of the results presented in this study might not be applicable in other contexts, potentially impacting the \emph{external validity} of our research. The questionnaire garnered 34 responses, 25 participants took part in our demo sessions and 11 in our interviews. These numbers, though in line with similar research conducted in industry~\cite{survey_response}, cannot be claimed to be representative of all types of development teams, application domains and other characteristics aside from those studied herein.
The conclusions derived from the interviews with Bloomberg software engineers might not resonate with software engineers from other organisations because of the different corporate cultures and development workflows employed at various organisations.

\section{Related Work}
\label{sec:relwork}
This work is relevant to research on automated program repair, deployments of program repair, user studies on program repair, recommender systems for software engineering, and code reviews.

\paragraph*{Automated program repair} APR aims to generate patches for software bugs automatically. A large body of research was dedicated to improving program repair tools~\cite{goues2019automated}, primarily focusing on improving their performance, such as scalability in Angelix~\cite{mechtaev2016angelix} and patch correctness in Prophet~\cite{long2016automatic}. A recent trend in improving performance characteristics has been incorporating machine learning as seen in Cure~\cite{jiang2021cure} and, in particular, large language models as in InferFix~\cite{jin2023inferfix}. However, the research on other important aspects of program repair, such as user experience, is limited. B-Assist attempts to fill this gap by adopting a user-centric approach.

\paragraph*{Deployments of program repair} Despite not always being a priority in previous research, several deployments of program repair tools have been successful. SapFix~\cite{marginean2019sapfix} was the first to demonstrate that end-to-end APR can work at scale in industrial practice. However, SapFix only repaired a narrow class of defects and did not focus on optimising user experience. Fixie~\cite{fixie} was deployed at Bloomberg and is currently available for use as one of the back-end APR tools (RSAT drivers) of B-Assist. Repairnator~\cite{urli2018design} is a program repair bot that generates patches for test failures in open-source projects on GitHub, reporting them through PRs. Contrary to Repairnator, B-Assist only provides suggestions in the context of existing PRs, placing a strong focus on suggesting relevant patches.

\paragraph*{User studies on program repair} The design of B-Assist was motivated by the previous study on APR conducted at Bloomberg~\cite{bloomberg_apr}. This study identified the importance of a user-centred approach to APR, which is characterised mainly by a strong focus on how and when fixes are presented to software engineers. B-Assist can be viewed as a specific approach to address the challenges outlined in that study. \citet{noller2022trust} conducted a large-scale study with software engineers evaluating their trust in APR. One of the findings from this study, being that software engineers would benefit from additional artefacts such as patch explanations, is consistent with our \Cref{finding:reason}.

\paragraph*{Recommender systems for software engineering} Research indicates that recommender systems for software engineering (RSSEs) are becoming increasingly important in contemporary coding practices~\cite{rsse}. A recent study~\cite{developers_using_multiple_tools} revealed that software engineers frequently employ a range of tools in their workflows, each addressing specific areas such as style, exceptions, and formatting. Review Bot~\cite{review_bot} consolidates the feedback from multiple tools, showcasing their corrections via a unified user interface. B-Assist can further enhance the developer experience by embedding code suggestions generated by APR tools directly into existing pull requests, eliminating the added layer of an external user interface that might divert attention during code reviews.

\paragraph*{Code review} A study on GitHub’s Suggested Changes~\cite{githubsuggestedchanges} showed that this feature helps support code review activities and improve collaboration between software engineers in the pull-based development model~\cite{chbrown_suggested_changes}. Software engineers have been proactive in adopting this feature, with more than 100K suggested changes created by one in ten reviewers within two weeks of its release. The study also found that the suggested change feature increased the amount of interaction between software engineers during code reviews. \citet{suggestion_bot} proposed the concept of integrating automated code review bots with GitHub’s Suggested Changes, providing a preliminary implementation of this in the form of Suggestion Bot. Further, \citet{change_level_suggestions} suggest through their research that “change-level” suggestions are best made early on whilst still fresh in software engineers’ minds. This point is further backed by prior research at Bloomberg, stating that the location and presentation of fixes are essential to consider when implementing new code analysis tools~\cite{bloomberg_apr}. Building on this prior work, B-Assist takes advantage of this feature as a means to realise its concept of presenting engineers with relevant and timely repairs.

\section{Discussion and Future Work}
\label{sec:discussion}
This section analyses the current limitations of B-Assist, discusses the impact of the tool and our study, as well as outlines directions for future research.

\paragraph*{Limitations} B-Assist has a number of technical limitations, such as the restrictions of GitHub's Suggested Changes format as discussed in \Cref{sec:limitations}, that reduce its effectiveness. We believe these limitations will be addressed with future releases of the underlying third-party components that B-Assist relies on.

Another limitation of B-Assist is that since it only generates patches in the context of existing PR changes, it may not be able to generate patch suggestions that span multiple locations or multiple files that are not all modified in the given PR. Some study participants confirmed this by expressing worries that filtering out lines that are not changed might produce incomplete fixes. We view this limitation as a trade-off between effectiveness and relevance. In future, this problem can be addressed by more intelligently defining the relevant scope of patch application.

Although patch suggestions provided by B-Assist are often relevant and timely, they may still be incorrect. This is because APR tools often generate patches that do not meet the software engineers' intentions but merely overfit the tests~\cite{smith2015cure} or analysed property~\cite{liu2023program}. We believe this problem will be addressed in future APR tools by reducing the rate of overfitting to an acceptable value. Aside from this concern, even an incorrect fix can be useful as it brings an issue to the engineers' attention, as noted in \Cref{sec:concluding_study_interview}.

\paragraph*{Impact} B-Assist will have a positive impact on Bloomberg's code review practices. The standard for collaborative bug-fixing at Bloomberg has been through creating separate pull requests containing required changes and merging these changes to the current working branch. One of the primary benefits provided by B-Assist is the streamlining of this process, allowing bug-fixing to be carried out directly within the current pull request through review comments. This new format removes a layer of abstraction that could be distracting for software engineers and contributes to a more lightweight code review process. It also makes patches generated by various tools more accessible at Bloomberg as, before introducing B-Assist, they were mostly unused by software engineers due to the lack of a convenient user interface.

Aside from demonstrating our approach's efficacy, this research emphasises the importance of user experience for APR. By pinpointing ongoing APR usability hurdles and showcasing a practical solution via B-Assist, we will catalyse future research to prioritise and innovate around user experience, setting a trajectory for the next generation of APR tools to be more developer-centric.

\paragraph*{Future work} Following our initial survey on B-Assist's concept, we were able to gather and act on the feedback by considering and implementing frequently requested features in our first prototype. For instance, two features that were not included in the original concept (namely, indicating the reason for a suggested fix and which APR tool generated it in the comment descriptions) were later implemented, even if the RSAT drivers currently available at Bloomberg are only able to provide only a very limited explanation for the patch generated and more work is needed in this respect in the future. An important avenue of future work for Bloomberg's user-centric APR would be to experiment with the automated generation of patch explanations, an increasingly realistic prospect, especially given the promising results from recent studies in this field with large language models~\cite{ssbse23ExplainableAPR}.

Given that B-Assist is language agnostic, we plan to extend its usage to other programming languages frequently used at Bloomberg, such as Java and Python. We will also provide a tighter integration with APR tools to provide more detailed and reliable information about defects and the corresponding fixes, which will be presented to software engineers alongside patch suggestions.

We also plan to measure the average time software engineers take to review and accept or reject patch suggestions, highlighting which aspects of our system affect this timing. This investigation would allow us to refine the tool and enable faster code reviews.

\section{Conclusion}
This study presents B-Assist, a tool developed and deployed at Bloomberg that adopts a user-centric approach to offer automatic patch suggestions for software defects. We emphasize the importance of the right timing, context, and audience to enhance the acceptance and usefulness of automated patches. B-Assist suggests patches within the context of existing pull requests, leading to higher acceptance rates, as confirmed by evaluations we conducted with Bloomberg software engineers. As the field advances, it is crucial to focus on the software engineers' experience and integrate automated program repair tools smoothly into regular development workflows. Our work sets the foundation for such advancements.


\section*{Acknowledgments}
This work has been supported by the ERC Advanced Grant no.741278 and the UKRI EPRSC Grant no.EP/P023991/1.
It has received ethical approval by Bloomberg and University College London (UCL/CSREC/T/91). 
We would like to thank all the participants to the user study and all the people -- Bloomberg software engineers, researchers, and the exceptional UCL alumni Dave Williams, Jaden Wan and Xiaohan Jiang Chen, who took part to this project via the UCL Industry Exchange Network (UCL IXN) -- who have worked with us towards this vision.

\bibliographystyle{ACM-Reference-Format}
\bibliography{bibliography}

\end{document}